\documentclass[11pt,twoside]{article}
\usepackage{asp2004}
\usepackage{psfig}
\usepackage{epsf}
\usepackage{graphics}
\usepackage{lscape}
\begin{document}
\title{NLTE Spectral Analysis of GW~Vir Pulsators}
\author{E. Reiff,$^1$ D. Jahn,$^1$ T. Rauch,$^{1}$ K. Werner,$^1$ and
  J. W. Kruk$^2$}
\affil{$^1$Institut f\"ur Astronomie und Astrophysik, Universit\"at
  T\"ubingen, Germany \\ 
$^2$Department of Physics and Astronomy, The Johns Hopkins University,
Baltimore, USA} 

\begin{abstract} 
GW~Vir variables are the pulsating members in the spectroscopic class
of the PG~1159 stars. In order to understand the characteristic
differences between pulsating and non-pulsating PG\,1159 stars, we 
analyse FUSE spectra of eleven objects, of which six are pulsating, by 
means of state-of-the-art NLTE model atmospheres. The numerous metal 
lines in the FUV spectra of these stars allow a precise determination 
of the photospheric parameters. We present here preliminary results of
our analysis.  
\end{abstract}

\section{Introduction}
\label{sec:introduction} 
GW~Vir variables \index{GW~Vir variables} belong to the spectroscopic 
class of the PG~1159 stars \index{PG~1159 stars} (Wesemael, Green \&
Liebert 1985), which is named after
the prototype \index{PG~1159$-$035} PG~1159$-$035. These objects are
strongly hydrogen-deficient post-AGB stars \index{post-AGB stars}
which pass through the hottest stage of stellar evolution. Their
effective temperatures range between 75~000 and 200~000\,K, surface
gravities vary from $\log g = 5.5-8.0\,\,[\mathrm{cm\,s}^{-2}]$. The
so-called born-again scenario (a late thermal pulse which 
transferred these objects back to the AGB followed by a second
post-AGB evolution, Iben et al\@. 1983) is mainly accepted as an
explanation for the H-deficiency and can reproduce well the observed
abundances. PG~1159 stars have spectra which are dominated by lines of
\ion{He}{ii}, \ion{C}{iv}, and \ion{O}{vi} (Werner et al. 2004), their
atmospheres show a typical surface composition of He:C:O = 33:50:17 by
mass. Beside these main constituents there are several lines of trace
elements, such as neon, nitrogen, silicon, sulfur, phosphorus, and
fluorine 
(Reiff et al. 2005, Werner et al. 2005). Presently 37 PG~1159 stars
are known, eleven of them proved to be pulsators. The pulsating members
of the PG~1159 class are referred as GW~Vir variables. They are
non-radial g-mode pulsators with periods from 300\,s up to 1000\,s, in
some cases exceeding even 2000\,s (Nagel \& Werner 2004). The favored
excitation mechanism for the pulsations is the
$\kappa$-mechanism associated with cyclic ionization of carbon and
oxygen (Quirion, Fontaine \& Brassard 2004). In the $\log
T_{\mathrm{eff}} - \log g$ diagram the GW~Vir variables are located
among the PG~1159 stars in the so-called GW~Vir instability
strip. Spectral analyses of pulsating and non-pulsating PG~1159 stars
were used by Dreizler \& Heber (1998) to define empirically the edges
of this instability strip. But it is still puzzling that also
non-pulsating PG~1159 stars are located within the instability
strip. In our analysis we try to find more characteristic properties
to distinguish between pulsating and non-pulsating members of this
class.  

\begin{figure}[t]
\centerline{\hbox{\psfig{figure=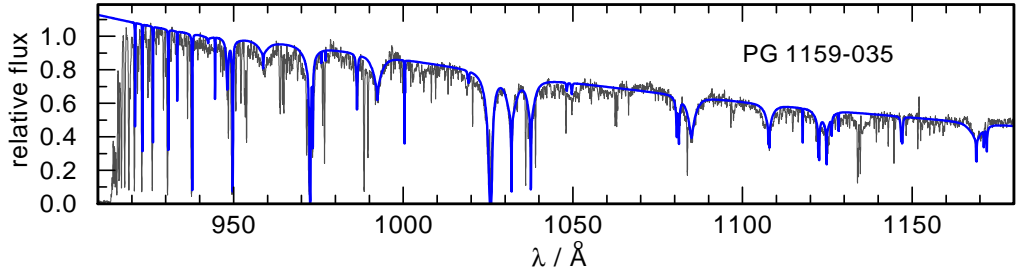,width=13.4cm}}}
\caption[]{FUSE\index{FUSE} spectrum and synthetic spectrum of 
  PG\,1159$-$035 (GW~Vir, $T_{\mathrm{eff}} = 140~000\,$K, $\log g = 7.0$).}     
\label{fig:pg1159} 
\end{figure}

\section{Observations and First Results}
\label{sec:observations_results}
For our analysis we selected pulsating and non-pulsating PG1159 stars
for which high resolution (R $\approx$ 20~000) FUV spectra obtained
with the Far Ultraviolet Spectroscopic Explorer (FUSE) are
available. The resulting sample comprises eleven objects. The FUSE
spectra are processed within the standard Calfuse pipeline process. A
log of all observations used for this analysis is listed in Table
\ref{tab:log}. Besides the FUV 
spectra we also used spectra obtained with STIS, GHRS and IUE as well
as optical spectra. The model atmospheres and synthetic 
line profiles are computed with the T\"ubingen Model Atmosphere Package 
(Werner et al\@. 2003, Rauch \& Deetjen 2003). The 
line-blanketed NLTE model atmospheres are in radiative and in   
hydrostatic equilibrium. Besides the main constituents of the
atmospheres of PG~1159 stars, helium, carbon, 
and oxygen, our model atmospheres also contain neon and 
nitrogen. For the abundances of these elements we use atmospheric
parameters taken from the literature which are summarized in Table
\ref{tab:parameters}. For neon an abundance of 2\% mass fraction was
assumed for all models, according to Werner \& Rauch (1994) and Werner
et al\@. (2004). Although the abundances in the literature were mostly 
determined in analyses of optical spectra the synthetic spectra can
also fit the FUV spectra well, which confirms the literature values
for
abundances in most cases. In Fig.\,\ref{fig:pg1159} we display the FUSE
spectrum of PG\,1159$-$035 together with our synthetic
spectrum. As lines of sulfur and silicon were also identified in
several objects we included those elements in the synthetic spectra,
too. Both were treated with line formation calculations without
back-reaction on the atmospheric structure. We assumed
solar abundances for both elements.

\begin{table}[t]
\caption{Log of the FUSE observations used for this analysis.}
\label{tab:log}
\footnotesize
\begin{center}
\begin{tabular}{l l r c}
\noalign{\smallskip}
\tableline
\noalign{\smallskip}
Object & Observation ID &\multicolumn{1}{c}{$t_\mathrm{exp}$} & Aperture \\
\noalign{\smallskip}
\tableline
\noalign{\smallskip}
RX\,J2117.1+3412\index{RX\,J2117.1+3412} & P1320501 & 8232\,s & LWRS\\
PG\,1144+005\index{PG\,1144+005}         & P1320201 & 6859\,s & LWRS\\ 
PG\,1520+525\index{PG\,1520+525}         & P1320101 & 3648\,s & LWRS\\ 
PG\,1159$-$035\index{PG\,1159-035}       & Q1090101 & 6321\,s & LWRS\\ 
K\,1$-$16\index{K\,1-16}                 & M1031010 &11271\,s & HIRS\\                
HS\,2324+3944 \index{HS\,2324+3944}      & P1320601 & 4004\,s & LWRS\\ 
Abell 78\index{Abell 78}                 & B1100101 & 9972\,s & LWRS\\ 
                                         & B1100102 & 7894\,s & LWRS\\ 
NGC 7094\index{NGC 7094}                 & P1043701 &23183\,s & LWRS\\
Abell 43\index{Abell 43}                 & B0520202 &12150\,s & LWRS\\         
PG\,1424+535\index{PG\,1424+535}         & P1320301 &11132\,s & LWRS\\ 
PG\,1707+427\index{PG\,1707+427}         & P1320401 &14599\,s & LWRS\\
\noalign{\smallskip}
\tableline
\end{tabular}
\end{center}
\end{table}

\begin{figure}[!hbt]
\centerline{\hbox{\psfig{figure=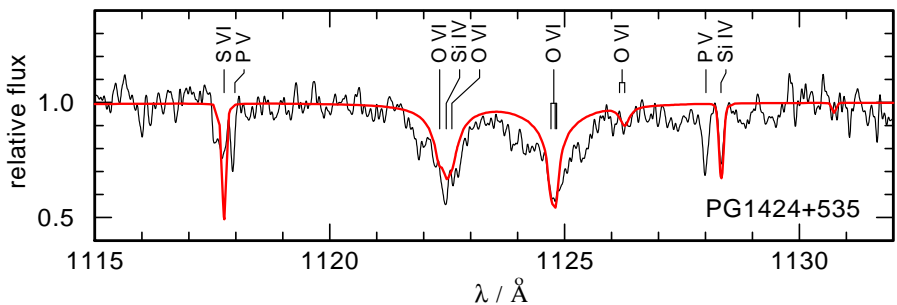,width=11.8cm}}}
\caption[]{Fit of the \ion{Si}{iv} doublet at 1122.5/1128.3\,\AA\ and
  the \ion{S}{vi} line at 1117.8\,\AA\ in the FUSE spectrum of 
  PG~1424+535. The \ion{P}{v} doublet at 1117.9/1128.0\,\AA\ is cleary 
  detected, but not yet included in our model.}      
\label{fig:SiS_lines} 
\end{figure}

Silicon is detectable in at least three objects, which are
PG~1159$-$035, and the two cooler stars PG~1424+535 and
PG~1707+427. Models with a solar Si abundance can fit the doublets at
1122/1128\,\AA\ and 1393/1402\,\AA\ well. In all spectra sulfur lines are
detected, but our preliminary fits also suggest abundances less than
solar. In Fig. \ref{fig:SiS_lines} we display part of the FUSE spectrum
of PG~1424+535 with a preliminary fit of the sulfur and silicon lines,
both with solar abundances.

\begin{figure}[!ht]
\centerline{\hbox{\psfig{figure=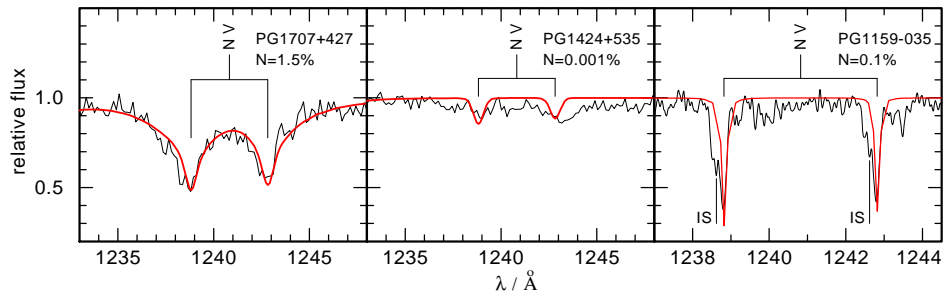,width=12.4cm}}}
\caption[]{Comparison of the N resonance doublet at
  1238.8/1242.8\,\AA\ in the GHRS spectra of PG~1707+427 
  ($T_{\mathrm{eff}} = 85~000\,$K, $\log g = 7.5$) and PG~1424+535 
  ($T_{\mathrm{eff}} = 110~000\,$K, $\log g = 7.0$) and in the STIS 
  spectrum of PG~1159$-$035 ($T_{\mathrm{eff}} = 140~000\,$K, $\log g
  = 7.0$) }     
\label{fig:N_comparison} 
\end{figure}

In former analyses by Dreizler \& Heber (1998) it was suggested that
the nitrogen abundance is a characteristic difference between
pulsating and non-pulsating PG~1159 stars, as nitrogen was detected in
all GW~Vir pulsators with a rather high abundance of 1\% by mass,
while in stable PG~1159 stars no nitrogen could be detected, except
for PG~1144+005 (which is considered outside the instability
strip). In order to confirm previously determined N abundances 
we tried to fit the N resonance doublet at 1238/1242\,\AA. For this
purpose we also analysed the STIS spectrum of PG~1159$-$035, which has
a high resolution (0.1\,\AA) and high S/N. In this spectrum the interstellar
component of the resonance doublet is clearly separated from the
photospheric component. This allows to determine the N abundance much
more precisely than before and it seems to turn out that the N abundance
is also significantly lower, about 0.1\% by mass, than suggested by
Dreizler \& Heber (1998). Fig.
\ref{fig:N_comparison} shows the N resonance doublets of three
objects, the pulsators PG~1159$-$035 and PG~1707+427 and the
non-pulsator PG~1424+535. While the comparison of PG~1707+427 and
PG~1424+535 seems to confirm the characteristic difference in the N
abundance, the new fit to the photospheric components in the STIS
spectrum of PG~1159$-$035 shows that the N abundance is only 0.1\% by
mass, but still higher than in the non-pulsator PG~1424+535. Further
analyses are necessary 
to confirm these preliminary results.

\begin{table}
\caption{Summary of the atmospheric parameters of our program stars
  taken from the literature. All objects are of spectral type PG1159
  except for Abell\,78, which is a [WC]-PG1159 transition object. The
  last column indicates whether the star is pulsating or not.} 
\label{tab:parameters}
\footnotesize
\begin{center}
\begin{tabular}{l r c r c r r r c}
\noalign{\smallskip}
\tableline
\noalign{\smallskip}
Object & $T_{\mathrm{eff}}$ & $\log g$ &\multicolumn{1}{c}{H} & He 
&\multicolumn{1}{c}{C} & \multicolumn{1}{c}{N} & \multicolumn{1}{c}{O} & Puls.\\
\noalign{\smallskip}
\cline{4-8} 
\noalign{\smallskip}
       & $[$kK$]$           & (cgs)  &\multicolumn{5}{c}{(mass fractions)} &\\
\noalign{\smallskip}
\tableline
\noalign{\smallskip}  
RX\,J2117.1+3412 & 170 & 6.0 &      & 38.0 & 56.0 &     &  6.0 & $\times$\\ 
PG\,1144+005     & 150 & 6.5 &      & 39.0 & 58.0 & 1.5 &  1.6 & \\                      
PG\,1520+525     & 150 & 7.5 &      & 44.0 & 39.0 &     & 17.0 & \\                    
PG\,1159$-$035   & 140 & 7.0 &      & 33.0 & 49.0 & 1.0 & 17.0 & $\times$\\ 
K 1$-$16         & 140 & 6.4 &      & 33.0 & 50.0 &     & 17.0 & $\times$\\  
HS\,2324+3944    & 130 & 6.2 & 21.0 & 41.0 & 37.0 &     &  1.0 & $\times$\\ 
Abell 78         & 110 & 5.5 &      & 33.0 & 50.0 & 2.0 & 15.0 & \\  
NGC 7094         & 110 & 5.7 & 36.0 & 43.0 & 21.0 &     &      & \\ 
Abell 43         & 110 & 5.7 & 36.0 & 43.0 & 21.0 &     &      & $\times$ \\                          
PG\,1424+535     & 110 & 7.0 &      & 50.0 & 44.0 &     &  6.0 & \\                    
PG\,1707+427     &  85 & 7.5 &      & 43.0 & 38.5 & 1.5 & 17.0 & $\times$\\ 
\noalign{\smallskip}
\tableline
\end{tabular}
\end{center}
\end{table}

\acknowledgements{This research is supported by the DFG under grant
  WE\,1312/30-1 (E.R.), by the DLR under grant 50\,OR\,0201 (T.R.) and
  the FUSE project, funded by NASA contract NAS5-32985 (J.W.K.).}

{}

\end{document}